\documentclass[sigconf]{acmart}

\usepackage{booktabs} 

\setcopyright{rightsretained}

\acmDOI{}

\acmISBN{}

\acmConference{Social Web in Emergency and Disaster Management}{2018}{Los Angeles, CA, USA} 

\copyrightyear{2018}

\begin{document}
\title{Geolocating social media posts for emergency mapping}

\subtitle{Demo paper}

\author{Barbara Pernici}
\affiliation{%
  \institution{Politecnico di Milano --- DEIB}
  \streetaddress{piazza Leonardo da Vinci 32}
  \city{Milano}
  \state{Italy} 
  \postcode{20133}
}
\email{barbara.pernici@polimi.it}

\author{Chiara Francalanci}
\affiliation{%
  \institution{Politecnico di Milano --- DEIB}
  \streetaddress{piazza Leonardo da Vinci 32}
  \city{Milano}
  \state{Italy} 
  \postcode{20133}
}
\email{chiara.francalanci@polimi.it}

\author{Gabriele Scalia}
\affiliation{%
  \institution{Politecnico di Milano --- DEIB}
  \streetaddress{piazza Leonardo da Vinci 32}
  \city{Milano}
  \state{Italy} 
  \postcode{20133}
}
\email{gabriele.scalia@polimi.it}

\author{Marco Corsi}
\affiliation{%
  \institution{e-Geos}
  \streetaddress{via Tiburtina, 965}
  \city{Roma}
  \state{Italy} 
  \postcode{00161}
}
\email{marco.corsi@e-geos.it}

\author{Domenico Grandoni}
\affiliation{%
  \institution{e-Geos}
  \streetaddress{via Tiburtina, 965}
  \city{Roma}
  \state{Italy} 
  \postcode{00161}
}
\email{domenico.grandoni@e-geos.it}

\author{Mariano Alfonso Biscardi}
\affiliation{%
  \institution{e-Geos}
  \streetaddress{via Tiburtina, 965}
  \city{Roma}
  \state{Italy} 
  \postcode{00161}
}
\email{mariano.biscardi@e-geos.it}

\renewcommand{\shortauthors}{B. Pernici et al.}

\begin{abstract}
The demo will illustrate the features of a webGIS interface to support the rapid mapping activities after a natural disaster, with the goal of providing additional information from social media to the mapping operators. This demo shows the first results of the E2mC H2020 European project, where the goal is to extract precisely located information from available social media sources, providing accurate geolocating functionalities and, starting from posts searched in Twitter, extending the social media exploration to Flickr, YouTube, and Instagram.
\end{abstract}

%
%


\maketitle

\section{Introduction}

\begin{figure*}
\centering
\includegraphics[width=0.8\textwidth]{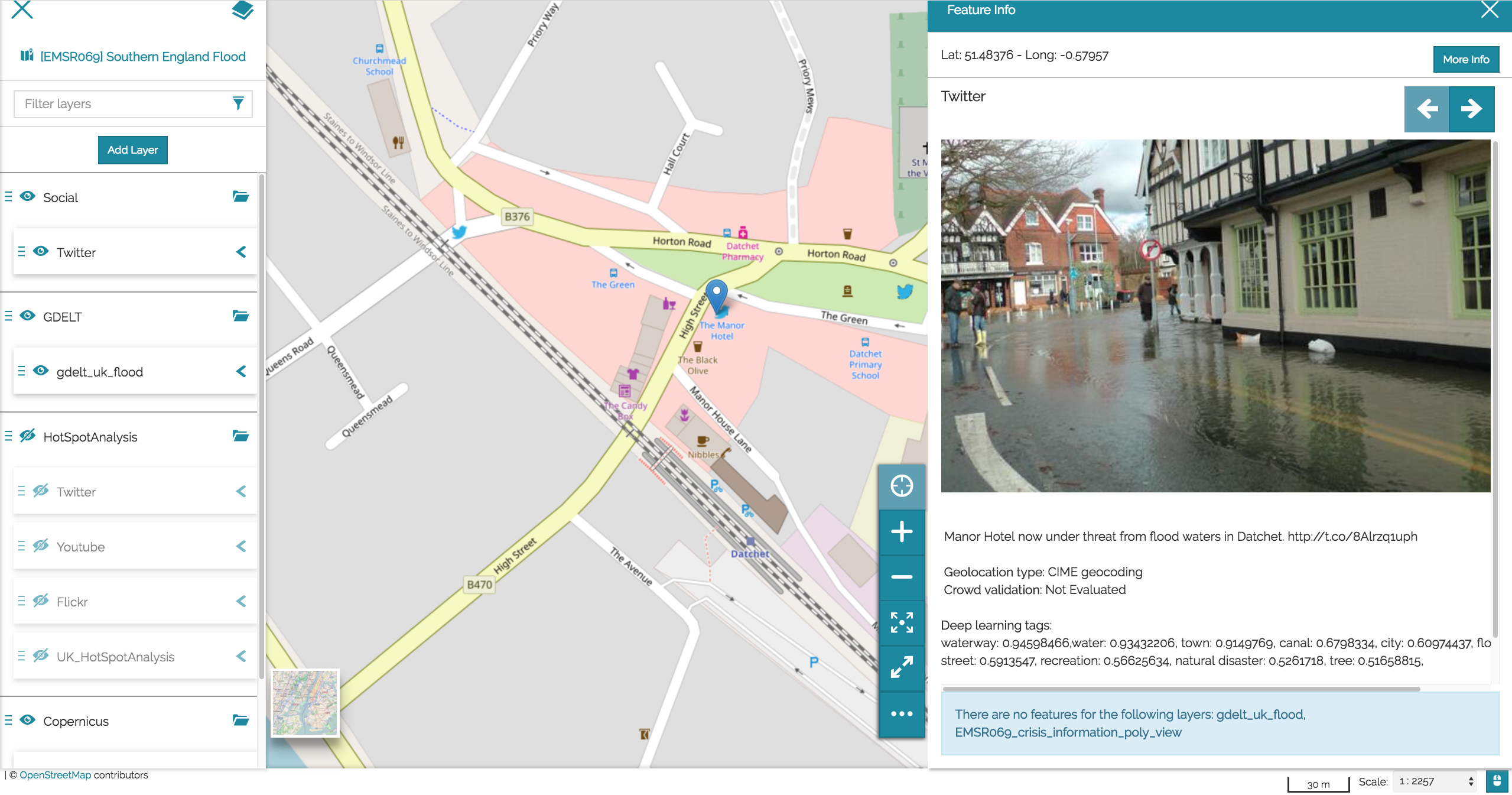}
\caption{Example of visualization in E2mC.}
\label{fig:webgis1}
\end{figure*}

In the E2mC (E2MC Evolution of Emergency Copernicus services) H2020 European project \cite{h2020,havas2017}, the goal is to provide additional information to operators working in rapid mapping activities based on satellite data in the Copernicus Emergency Rapid Mapping Service, based on satellite data (\url{http://emergency.copernicus.eu/mapping}). Rapid mapping has the goal of providing rescue teams and operators with information about the current situation of the area being interested by the emergency. The information has to be provided in a rapid, systematic, and organized way, with the main goal of making mapping faster. 

The extraction of information from social media has been studied by several authors in the literature, in particular in emergency and crisis situations \cite{castillo2016big,imran2015processing}, also with ad-hoc initiatives\footnote{Some examples include Humanitarian OpenStreetMap (\url{https://www.hotosm.org/}), CrisisMappers (\url{http://crisismappers.net/}) and Facebook Disaster Maps (\url{https://research.fb.com/facebook-disaster-maps-methodology/}).}. In this paper we focus on extracting visual information from social media, that can provide a more precise and timely view of disaster areas. As discussed in the literature, in which tweets have been studied extensively, one of the limitations is that only a limited number of tweets is natively georeferenced (or geotagged), and in general the numbers reported in the literature indicate 3\% or less of the tweets. Moreover, the georeferenced tweets are not necessarily related to the occurring events, although it has been reported in \cite{DBLP:journals/gis/AlbuquerqueHBZ15} that in the areas of interest of the emergency the probability of the tweet being related to the emergency is much higher than in other more distant areas. In addition, for being valuable in supporting the operators, the locations of the tweets needs to be precise. While most geolocation algorithms focus on identifying locations (localities or points of interests), only recently the challenges related to precise location identification have been addressed  \cite{DBLP:journals/snam/Paraskevopoulos16}. Another aspect of previous work is the focus on the text of the tweets, while other information of tweets is rarely considered.

On this basis, we have developed a context-based geolocation framework focusing on image extraction, called CIME \cite{cime2017}. The system is based on  multilingual Stanford CoreNLP \cite{DBLP:conf/acl/ManningSBFBM14} for Named Entity Resolution and OpenStreetMap \cite{DBLP:journals/pervasive/HaklayW08} as a basis for identifying locations. CIME uses both the local context of the tweet (text and metadata associated to the post) and its global context (relations to other tweets in terms of commons hashtags, retweets, mentions, and the like). In addition, images and videos are extracted not only directly from the tweets, but also following linked social media.

\section{Demo scenario}
The demo will illustrate the presentation of the results of the image extraction activity. The visualization follows an explorative approach to the available information which is provided to the operators.

The main functionalities which will be illustrated in the demo are the following:
\begin{itemize}
\item Visualization based on spatial and temporal selection of the area and time of interest.
\item Separate and joint visualization of georeferenced tweets, and of those localized with the CIME algorithm.
\item Visualization of geolocated media contained in the tweets, and also media linked by the tweets.
\item The possibility of selecting tweets using a ranking function, which depends on the precision of the tweet geolocation, and of modifying the visualization options and ranking by the operator.
\end{itemize}

Fig. \ref{fig:webgis1} shows an example of a geolocalized tweet visualization, based on the 2014 UK floods. The platform shows the information associated to the tweet (text and media), with a link to the original post, and information associated to its analysis: geolocation type, if it has been manually validated by the crowd and tags obtained analyzing the image. Moreover, a link to the location in Google Street View\footnote{\url{https://www.google.it/streetview/}} is provided. Other case studies (Central Italy earthquake in 2016, Harvey storm in 2017) will also be demonstrated.

\section{System characteristics}
The E2mC visualizer is a web application, interacting with a geographic web server GeoServer 2.9.4, using a WMS OGC standard service for layering. Post IDs and location information are stored in a PostgreSQL 9.6 database with the spatial extension PostGIS. 

\section{Concluding remarks}
The E2mC project will further develop the explorative approach to retrieve social media information related to an emergency. In addition to the basic location layer for tweets, additional information will be provided such as analyzing hotspots, developing image analysis tools to compare and classify images, multilingual support for topic extraction, and crowdsourcing functionalities, which are going to be integrated in the continuation of the project as described in \cite{havas2017}.

\begin{acks}
This work has been partially funded by the European Commission H2020 project E2mC ``Evolution of Emergency Copernicus services'' under project No. 730082. This work expresses the opinions of the authors and not necessarily those of the European Commission. The European Commission is not liable for any use that may be made of the information contained in this work. 
\end{acks}

\bibliographystyle{abbrv}
\bibliography{biblio} 

\begin{thebibliography}{1}

\bibitem{h2020}
{H2020 project E2MC Evolution of Emergency Copernicus services
  https://www.e2mc-project.eu/}.

\bibitem{castillo2016big}
C.~Castillo.
\newblock {\em Big Crisis Data: Social Media in Disasters and Time-Critical
  Situations}.
\newblock Cambridge University Press, 2016.

\bibitem{DBLP:journals/gis/AlbuquerqueHBZ15}
J.~P. de~Albuquerque, B.~Herfort, A.~Brenning, and A.~Zipf.
\newblock A geographic approach for combining social media and authoritative
  data towards identifying useful information for disaster management.
\newblock {\em International Journal of Geographical Information Science},
  29(4):667--689, 2015.

\bibitem{DBLP:journals/pervasive/HaklayW08}
M.~M. Haklay and P.~Weber.
\newblock {OpenStreetMap}: {U}ser-generated street maps.
\newblock {\em {IEEE} Pervasive Computing}, 7(4):12--18, 2008.

\bibitem{havas2017}
C.~Havas, B.~Resch, C.~Francalanci, B.~Pernici, G.~Scalia, J.~L.
  Fernandez-Marquez, T.~V. Achte, G.~Zeug, R.~Mondardini, D.~Grandoni,
  B.~Kirsch, M.~Kalas, V.~Lorini, and S.~R{\"{u}}ping.
\newblock E2mc: Improving emergency management service practice through social
  media and crowdsourcing analysis in near real time.
\newblock {\em Sensors}, 17(12), 2017.

\bibitem{imran2015processing}
M.~Imran, C.~Castillo, F.~Diaz, and S.~Vieweg.
\newblock Processing social media messages in mass emergency: A survey.
\newblock {\em ACM Computing Surveys (CSUR)}, 47(4):67, 2015.

\bibitem{DBLP:conf/acl/ManningSBFBM14}
C.~D. Manning, M.~Surdeanu, J.~Bauer, J.~R. Finkel, S.~Bethard, and
  D.~McClosky.
\newblock The {Stanford CoreNLP} natural language processing toolkit.
\newblock In {\em Proceedings of the 52nd Annual Meeting of the Association for
  Computational Linguistics, {ACL} 2014, June 22-27, 2014, Baltimore, MD, USA,
  System Demonstrations}, pages 55--60, 2014.

\bibitem{DBLP:journals/snam/Paraskevopoulos16}
P.~Paraskevopoulos and T.~Palpanas.
\newblock Where has this tweet come from? {Fast} and fine-grained
  geolocalization of non-geotagged tweets.
\newblock {\em Social Netw. Analys. Mining}, 6(1):89:1--89:16, 2016.

\bibitem{cime2017}
G.~Scalia, C.~Francalanci, and B.~Pernici.
\newblock Cime: Social context for image geolocalization.
\newblock {\em Submitted for publication}.

\end{thebibliography}

\end{document}